\documentclass[final,5p,times,twocolumn]{elsarticle}

\usepackage{amssymb,amsmath}
\usepackage{textcomp}
\usepackage{subfigure}
\usepackage{braket}
\usepackage{float}
\usepackage{xcolor}
\usepackage{cleveref}
\usepackage{multirow}
\usepackage{icomma}
\usepackage{stackrel}
\usepackage{appendix}

\begin{document}

\title{Dynamics in fractal spaces}

\author{Eugenio Meg\'{\i}as}
 \address{Departamento de Física Atómica, Molecular y Nuclear and Instituto Carlos I de Física Teórica y Computacional, \\
 Universidad de Granada, Avenida de Fuente Nueva s/n, 18071 Granada, Spain
 }
\ead{emegias@ugr.es}

\author{Alireza Khalili Golmankhaneh}
 \address{Department of Physics, Urmia Branch, Islamic Azad University, Urmia, Iran
 }
\ead{ealirezakhalili2005@gmail.com}

\author{Airton Deppman}
 \address{Instituto de Física -  Universidade de São Paulo, São Paulo 05508-090, Brazil
 }
\ead{deppman@usp.br}

%


\begin{abstract}
This study investigates the interconnections between the traditional Fokker-Planck Equation (FPE) and its fractal counterpart (FFPE), utilizing fractal derivatives. By examining the continuous approximation of fractal derivatives in the FPE, it derives the Plastino-Plastino Equation (PPE), which is commonly associated with Tsallis Statistics. This work deduces the connections between the entropic index and the geometric quantities related to the fractal dimension. Furthermore, it analyzes the implications of these relationships on the dynamics of systems in fractal spaces. In order to assess the effectiveness of both equations, numerical solutions are compared within the context of complex systems dynamics, specifically examining the behaviours of quark-gluon plasma (QGP). The FFPE provides an appropriate description of the dynamics of fractal systems by accounting for the fractal nature of the momentum space, exhibiting distinct behaviours compared to the traditional FPE due to the system's fractal nature. The findings indicate that the fractal equation and its continuous approximation yield similar results in studying dynamics, thereby allowing for interchangeability based on the specific problem at hand.
\end{abstract}


\maketitle

\section{Introduction}

Since the pioneering work by Haussdorff, more than a century ago, the field of complex geometrical spaces with fractional dimensions has flourished as one of the most active areas of scientific investigation. Today objects with fractional dimensions are known as fractals, a term coined by Mandelbrot~\cite{Mandelbrot1982-jg}. The fundamentals of the mathematical aspects related to those systems can be found in Ref.~\cite{Falconer2014-de}.

The advances in the formal aspects of fractals have evolved to include Calculus and several possible generalizations of the standard derivative have been considered, such as the fractional derivatives~\cite{Li2007-oq,Abdeljawad2011-lt} and the fractal derivatives~\cite{He2014-ng,He2018-zd,Golmankhaneh2016-hq}. Many of the proposed derivative forms have problems being defined in a general way~\cite{Valerio2022-ia}. This work will use the fractal derivative proposed by Parvate and Gangal~\cite{FD1,FD2}, which is rooted in Haussdorff's measure and dimension. A comprehensive account of fractal calculus can be found in Ref.~\cite{Golmankhaneh2022-vn,golmankhaneh2021tsallis}, where the $q$-fractal calculus was introduced to derive $q$-Gaussian, generalized stable, and the L\'{e}vy distributions on fractal sets.

One important consequence of the nonextensivity of systems with non-local correlations is the fact that the collisional term of the Boltzmann Equation (BE) is modified, with the correlation functional assuming a different form~\cite{Lima:2001lgd,Lavagno:2002hv}. In Ref.~\cite{Deppman:2023ztz}, it was shown that this new functional form leads to the Plastino-Plastino Equation (PPE)~\cite{PLASTINO1995347}, which was previously introduced as the framework for dealing with nonextensive systems, in substitution to the Fokker-Planck Equation (FPE), and has been applied in many studies~\cite{Andrade2010-ve,O_Manipon2023-ej,Lutz2013-zz,Combe2015-hc,Gomez2023-lm}. In this context, it is important to point out that a fractional derivative version of the FPE (FFPE) was proposed in Ref.~\cite{Golmankhaneh2016-hq}.

The statistical aspects of fractal systems have often been associated with Tsallis' nonextensive  statistics~\cite{Tsallis,TsallisBook}. In this context, studies of anomalous diffusion have benefitted from the generalized statistics. The multiparticle production occurring in high-energy collisions, as well, is a field where the nonextensive statistics has an important impact. Thermofractals are systems that have fractal structures in the thermodynamical distribution~\cite{Deppman-PRD-2016,DMM-Frederico-2018}, and they are described by the nonextensive statistics. This kind of structure explains the emergence of $q$-exponential distributions in multiparticle production at high energy collisions~\cite{DMM-PRD-2020}.

The present work investigates the dynamic equation for diffusion in fractal spaces. It introduces fractal derivatives into the Fokker-Planck Equation and employs a continuous approximation to derive the Plastino-Plastino Equation, which is commonly associated with systems demonstrating non-local correlations and non-extensive statistics in their dynamics. The relationships between the quantities characterizing fractal geometry and the entropic index are determined. As an illustrative application, the study explores the intriguing problem of a heavy-quark motion within the Quark-Gluon Plasma. Both the fractal version of the Fokker-Planck Equation and the Plastino-Plastino Equation are used to investigate this problem, and the results demonstrate the effectiveness of the formalism developed here.

The Fokker-Planck Equation (FPE) is an important tool for investigating the dynamic properties of complex systems. It has applications in Astrophysics, Cosmology, hydrodynamics, and granular dynamics. The dynamical aspects of the quarks in a hot and/or dense quark medium have been of intense investigation since they can give information on the properties of the QGP~\cite{Svetitsky:1987gq,WaltonRafelski}. Aside from the QGP, the study of neutron star properties is another field which can benefit from a deeper understanding of the quark and gluon dynamics. This work investigates the relationships between the fractal derivatives and the nonextensive distributions. In particular, we show that the "special derivative" used in Ref.~\cite{Deppman:2023ztz} to derive the PPE from the BE for systems with non-local correlation can be described as the inverse of the fractal derivative. Then we describe the FFPE and compare it with the PPE.

This work is organized as follows: Section 2 describes the fractal version of the Fokker-Planck Equation and applies the continuous approximation to obtain the Plastino-Plastino Equation, relating the relevant fractal geometry quantities to the entropic parameter in the PPE. Section 3 presents an analysis of the FFPE and PPE equations, and finds the relationship between the geometrical quantities and the entropic index, showing that they are identical to those obtained previously on the basis of the fractal geometry. Section 4 uses the problem of the heavy-quark dynamics in the QGP to show how the developed formalism works in practice. The results are discussed and compared aiming to verify the consistency of the formalism developed in the previous sections.

\section{Fractal derivatives and Fokker-Planck equation}

A fundamental quantity in Haussdorff's approach to geometries with fractional dimension is the mass distribution, $\gamma(\mathbb{F},a,b)$, where $\mathbb{F}$ is a fractal space, and $a,b \in \mathbb{F}$. Based on the mass distribution, the staircase function is defined by
\begin{equation}
    S^{\alpha}_{F,a_o}=\begin{cases}
        \gamma(\mathbb{F},a_o,x) ~~\textrm{for $x > a_o$} \\
        \gamma(\mathbb{F},x,a_o) ~~\textrm{for $x < a_o$}
    \end{cases}\,, \label{staircase}
\end{equation}
where $a_o \in \mathbb{F}$.

Let $\mathbb{F}$ be a fractal space embedded in $\mathbb{R}$  and $S^{\alpha}_{\mathbb{F},a_o}(x)$ be the $\alpha$-dimensional staircase function at $x$. The fractal derivative is defined as
\begin{equation}
    D_{\mathbb{F},a_o}^{\alpha}f(x_o)=\begin{cases}
    F\lim_{x \rightarrow x_o}\frac{f(x)-f(x_o)}{S_{F,a_o}^{\alpha}(x)-S_{F,a_o}^{\alpha}(x_o)} ~~~~x, x_o\in \mathbb{F} \\
    0 ~~~~\textrm{otherwise}
    \end{cases}\,, \label{eq:Dalpha}
\end{equation}
for any function $f:\mathbb{F} \rightarrow \mathbb{R}$. 

Ref.~\cite{DMP2023} extends the fractal derivative on fractal space, where the domain space is fractal, to include a derivative of fractal functions, where the image space is fractal. In this case, it results
\begin{equation}
[D_{\mathbb{F},\varphi}^{\alpha}]^{-1}h(x_o)=F\lim_{x \rightarrow x_o}\frac{S_{F,\varphi}^{\alpha}[h(x)]-S_{F,\varphi}^{\alpha}[h(x_o)]}{x-x_o}\,,
\end{equation}
where $\varphi=f(a_o)$ has the same role as $a_o$ in the staircase function. Observe that in this case, the image space and the domain space of the function $h$, are the same, i.e., function $h:\mathbb{F} \rightarrow \mathbb{F}$.
The fractal derivative is well defined for any pair $x, x_d  \in \mathbb{F}$, so it results in a fractal function with the same dimension as the function $f$.

The same work introduces the continuous approximation for the staircase differential as
\begin{equation}
    dS_{F,a_o}^{\alpha}(x)=\frac{A(\alpha)}{\alpha}dx^{\alpha}\,,
\end{equation} 
where $A(\alpha) := 2 \pi^{\alpha/2} / \Gamma(\alpha/2)\,$, with $\Gamma(x)$ being the Euler Gamma Function.  If $x$ or $x+dx$ are not in the fractal space $\mathbb{F}$, then $dS_F^{\alpha}(x)=0$, so in this case the continuous approximation is such that $dx^{\alpha}=0$. Using the equations above in the equation for the inverse of the fractal derivative, we note that the space where the derivative is applied is the image space of the fractal function $h(x)$, so we have $dS_{F,\varphi}^{\alpha}[h(x)] = (A(\alpha)/\alpha) dh^{\alpha}(x)$ and then
\begin{equation}
    [D_{\mathbb{F},\varphi}^{\alpha}]^{-1} h(x) = \frac{A(\alpha)}{\alpha}  \frac{dh^{\alpha}\left(x\right)}{dx}=A(\alpha)h^{\alpha-1}(x) \frac{dh}{dx}(x) \,. \label{InverseContinuousDerivative}
\end{equation}
The derivative in the equation above can be associated with the $q$-deformed calculus, introduced in Ref.~\cite{Borges-qCalculus}, and also with some important versions of fractional derivatives.

The work in Ref.~\cite{Deppman:2023ztz} used the $q$-derivative to derive the Plastino-Plastino Equation (PPE) from the Boltzmann Equation for systems strongly correlated. This investigation elucidated a potential correlation between fractal geometry and the equation governing non-linear dynamics, thereby substantiating the motivation for a comprehensive comparative analysis of both the conventional PPE and its fractal variant.

An application of the formalism discussed above is the Fractal Fokker-Planck Equation (FFPE). It generalizes the traditional Fokker-Planck Equation (FPE), which is given by
\begin{equation}
 \frac{\partial h}{\partial t}- \frac{\partial}{\partial p_i} \left[A_i(\mathbf p) h+ \frac{\partial }{\partial p_j} \left( B_{ij}(\mathbf p) h \right) \right] = 0 \,, \label{FPE}
\end{equation}
by using the fractal derivatives~\cite{Golmankhaneh2016-hq}, resulting in
\begin{equation}
  [D_{F',t}^{\beta}]h(\mathbf{p},t)= [D_{F,p_{o,i}}^{\alpha}]\left(A_i h(\mathbf{p},t)+ [D_{F,p_{0,j}}^{\alpha}](B_{ij} h(\mathbf{p},t) ) \right) \,. \label{IFFPE}
\end{equation}

Notice that the fractal dimensions for the momentum ($\alpha$) and time ($\beta$) spaces can be different. This aspect brings interesting consequences since there are many possibilities for addressing the effects of fractal spaces in dynamical systems, i.e. if the fractal dimension is a characteristic aspect of the system and must be the same for the function and its derivatives, the parameter $q$ must be the same for the function and its derivatives, as well.

In Ref.~\cite{DMP2023} it was argued that a fractal function and its derivative have the same dimensions. The continuous approximation should, conveniently, present the same property, which in this case manifests itself in the same parameter $q$ for both the function and its derivative. This property is present also in those systems governed by the non-additive entropy, as explained in Ref.~\cite{Deppman:2023ztz}, where it was used to derive the Plastino-Plastino Equation. The essential aspect, present also in the $q$-deformed calculus, is that
\begin{equation}
    h^{1-q}\frac{\partial h}{\partial x} \propto \frac{\partial h^{2-q}}{\partial x}\,.
\end{equation}
Comparing the equation above with Eq.~(\ref{InverseContinuousDerivative}), it follows that the entropic index, $q$, present in the PPE can be related to the fractal dimension $\alpha$, present in the fractal derivative, by the equation
\begin{equation}
    1-\alpha=q-1 \,, \label{eq:alpha-q-1D}
\end{equation} 
assuming $\alpha<1$.  The possibility of relating the entropic index and the fractal dimension indicates the deep relationship between Tsallis Statistics and Fractal Geometry, and it deserves to be investigated further.

The fractal space that is represented in Eq.~(\ref{eq:alpha-q-1D}) by its fractal dimension $\alpha$ is the image space of the distribution $f$ that appears in the PPE. 
Up to now, the discussion is restricted to a single spatial dimension, but it is possible to advance a geometrical interpretation for Eq.~(\ref{eq:alpha-q-1D}) by identifying $1-\alpha$ with the fraction of the space dimension that is removed to form the fractal space with dimension $\alpha=d_f$. The result can be generalized to any fractal dimension, by introducing the symbol $\lceil x \rceil$ to represent the integer such that
\begin{equation}
    \lceil x \rceil -1< x < \lceil x \rceil \,,
\end{equation}
and defining the dimension $d= \lceil d_f \rceil$. Then, we have
\begin{equation}
    q-1=\frac{d-d_f}{d} \,.  \label{eq:q1ddf}
\end{equation}
The geometrical significance of $q-1$ is kept and, when $d>1$, instead of a fractal function one has a fractal vector field~\cite{DMP2023}. 

The results obtained above can be tested by evaluating the effects of Eq.~(\ref{eq:q1ddf}) in the observable aspects of the fractal. This work shows that the dynamic aspects of a fractal, as described by a fractal derivative version of the Fokker-Planck Equation, agree with that assumption. It will be convenient to introduce the fractal dimension gap, $\delta d_f$, defined as
\begin{equation}
    \delta d_f \equiv d-d_f\,. \label{eq:fractalgap}
\end{equation}

The results obtained above are useful to relate the PPE with the fractal derivative of a fractal function, that is when the image space is fractal, but not necessarily the function domain. However, works have been done relating the dynamical equations to the fractal dynamical equations when the domain is a fractal space. It is useful to find a method to relate the parameter $q$ and the image fractal dimension $\alpha$ with the dimension in the domain space, in order to have the same dynamical properties. In other words, what is the dimension of a fractal space necessary to produce the same fractal properties described in the dynamical equation? 

In Eq.~(\ref{IFFPE}), the Fokker-Planck Equation was written in terms of the fractal derivatives, where the momentum and the time are considered as fractal spaces. According to the methodology proposed above, it is convenient to express the fractal gap in terms of the whole fractal dimension of the joint space formed by momentum and time, so that the fractal dimension is $d_F \ge d_{Ft}+d_{Fp}$~\cite{Falconer2014-de}, with $d_{Ft}$ and $d_{Fp}$ being, respectively, the time and momentum fractal dimensions. With the dimension $d_F$ and Eq.~(\ref{eq:fractalgap}), one can write Eq.~(\ref{IFFPE}) as
\begin{equation}
  [D_{F',t}^{\zeta}]h(\mathbf{p},t)= [D_{F,p_{o,i}}^{\zeta}]\left(A_i h(\mathbf{p},t)+ [D_{F,p_{0,j}}^{\zeta}](B_{ij} h(\mathbf{p},t) ) \right) \,, \label{IIFFPE}
\end{equation}
where we have defined $\zeta \equiv 2 - q$. The whole fractal dimension fraction contains all the information about the fractal.

Using the equation above and the continuous approximation in Eq.~(\ref{IFFPE}), one obtains the continuous approximation for the fractal FPE as
\begin{equation}
 \frac{\partial h}{\partial t}= \frac{\partial}{\partial p_i} \left[A_i(\mathbf p) h+ \frac{\partial }{\partial p_j} \left( B_{ij}(\mathbf p) h^{2-q_{\zeta}} \right) \right]  \,, \label{pPPE}
\end{equation}
where $q_{\zeta}=2-\zeta$ \label{pPPE-2},
which is the Plastino-Plastino Equation (PPE). This result shows that the PPE is the continuous approximation of the fractal version of the FPE.

The next parts of the present work will make a comparative study of the solutions of each of those equations associated with the dynamic evolution of complex systems. Understanding the similarities and differences among the solutions for these equations can help in finding the possible connections between the non-additive equation and the fractal equation. Despite the use of QGP physics to make this comparison, the results that will be obtained here are of general interest.

\section{Dynamics in a fractal medium}

Fractional derivatives, such as those derived in the preceding section through a continuous approximation of a fractal function, find numerous applications in the examination of dynamics within non-linear media. In this section, their utility is extended to dynamic systems described by the Fokker-Planck Equation. While the following discussion remains quite general, for the sake of clarity and the provision of numerical examples the present work applies these concepts to the specific scenario of a heavy-quark moving in the quark-gluon plasma. This problem is well-established and has been previously addressed using both the conventional Fokker-Planck Equation and the Plastino-Plastino Equation~\cite{Megias2023}. For the microscopic determination of transport coefficients, we adhere to the methodology outlined in Ref.~\cite{Svetitsky:1987gq}.

\subsection{PPE in a $d$-dimensional space}

In this section, a detailed investigation of the solutions for both the PPE and the fractal FPE (FFPE) will be undertaken. The objective is to find, from a dynamical point of view, the relation between the parameter $q$ in the PPE and the fractal space dimension in the FFPE. It will be shown that Eq.~(\ref{eq:q1ddf}) is exactly the relation needed to get the correct dynamical description of the dynamical behaviour in a fractal space.

The PPE in $d$ spatial dimensions is given by
\begin{equation}
 \frac{\partial f}{\partial t} = \sum_{i=1}^d \frac{\partial}{\partial p_i} \left[A_i(p^2) f + \sum_{j=1}^d \frac{\partial }{\partial p_j} \left( B_{ij}(p^2) f^{2-q} \right) \right]  \,.\label{pPPE_d}
\end{equation}
The transport coefficients may be decomposed as
\begin{eqnarray}
A_i({\mathbf p}) &=& p_i \, A(p^2) \,, \label{eqA}\\
B_{ij}({\mathbf p}) &=& \left( \delta_{ij} - \frac{p_i p_j}{p^2} \right) B_0(p^2) + \frac{p_i p_j}{p^2} B_1(p^2) \,. \label{eqB}
\end{eqnarray}
The coefficients $B_0(p^2)$ and $B_1(p^2)$ differ by a negligible amount, so we use the approximation $B_0(p^2)=B_1(p^2) \equiv B(p^2)$, and in this case we have $B_{ij}({\mathbf p}) =\delta_{ij}B(p^2)$.

The transport coefficients $A(p^2)$ and $B(p^2)$ are slowly increasing with $p$ in the relevant region of the present study, so we approximate both to constants $A$ and $B$, respectively. These approximations were discussed in some detail in Ref.~\cite{WaltonRafelski}.
A thorough discussion of the solutions of the PPE can be found in Ref.~\cite{Deppman:2023ztz}, and a comparison with the FPE solutions was presented in Ref.~\cite{Megias2023}. The mais objective of the present work is to compare the PPE and the FFPE solutions and investigate the role of the fractal dimension in the dynamics of the system in a fractal medium.

This equation has the following solution
\begin{equation}
 f(\mathbf p,t)=  \frac{\lambda^d}{\left(\sqrt{2\pi\chi_q}  \sigma(t) \right)^d} ~~\exp_q\left[-\frac{(\mathbf p - \mathbf p_M(t))^2}{2 \sigma(t)^2} \right] \,, \label{PPEsol-2}
\end{equation}
where $\lambda$ is a multiplicative factor which has dimensions of energy, and the parameters $\mathbf p_M(t)$ and $\sigma(t)$ write 
\begin{equation}
 \begin{cases}
  \mathbf p_{M}(t)=\mathbf p_{o} \exp[-A t] \\
  \\
\sigma(t) = \sigma_o\left[ \left(1 - \kappa  \right) 
\exp\left[-\xi At \right]  + \kappa \right]^{\frac{1}\xi} 
 \end{cases}\,,
 \label{soltutionparameters_d}
\end{equation}
with
\begin{equation}
    \kappa \equiv (2-q) \left( 2\pi \chi_q\right)^{\frac{d}{2}(q-1)} \frac{B}{A \sigma_o^2}  \left( \frac{\sigma_o}{\lambda} \right)^{d(q-1)} \,,
\end{equation}
and
\begin{equation}
\chi_q \equiv \frac{1}{q-1} \left( \frac{\Gamma\left( \frac{1}{q-1} - \frac{d}{2}\right) }{\Gamma\left( \frac{1}{q-1} \right)} \right)^{\frac{2}{d}}   \qquad q > 1 \,. \label{eq:chiq_d}
\end{equation}
In the equations above, the quantity $\xi$ is
\begin{equation}
\xi = 2 + d(1-q) = 2 - \delta d_f \,,
\end{equation}
where Eqs.~(\ref{eq:q1ddf}) and (\ref{eq:fractalgap}) have been used. 
The distribution is normalized as
\begin{equation}
\int d^dp \, f(\mathbf p, t) = \lambda^d \,. \label{eq:norm_d}
\end{equation}

From the solution of the PPE, it is clear that the fractal dimension gap $\delta d_f$ is fundamental in the determination of the distribution width at each and every time. This is reasonable if one takes into account that the medium porosity has strong effects in the diffusion process. When the fractal dimension vanishes, so that $\delta d_f \rightarrow 0$, $q \rightarrow 1$ and $\xi \rightarrow 2$, it is recovered the values observed in the FPE solutions. In this case, the fractal solution and the PPE solution becomes identical to the standard Fokker-Planck Equation solution for any dimension $d$.

\subsection{FFPE in a $d$-dimensional space}

The next step is the investigation of the FFPE solutions in $d$~spatial dimensions. The equation follows directly form Eq.~(\ref{IIFFPE}), and is given by
\begin{equation}
  [D_{F',t}^{\beta}]f(\mathbf{p},t)= \sum_{i=1}^d [D_{F,p_{o,i}}^{\alpha}]\left(A_i f(\mathbf{p},t)+ \sum_{j=1}^d [D_{F,p_{0,j}}^{\alpha}](B_{ij} f(\mathbf{p},t) ) \right) \,. \label{IFFPE_d}
\end{equation}

The general solution for the equation above is
\begin{equation}
  f(\mathbf{p},t) = \frac{ (\lambda \alpha)^d}{\left(\sqrt{2} \sigma_F(t)\right)^{d/\alpha} \Gamma\left[ \frac{1}{2\alpha} \right]^d}\exp\left(-\frac{(\mathbf{p}^{\alpha}-\mathbf{p}^{\alpha}_{M}(t))^{2}}{2\sigma_F(t)^{2}}\right) \,, \label{eq:f_fractal-2}
\end{equation}
where 
\begin{align}
p^{\alpha}&\propto \left( \sum_{i=1}^d p_{i}^{2\alpha} \right)^{1/2} \,, \\
 \mathbf{p}_{M}^{\alpha}(t)&\propto
\mathbf{p}_{o}^{\alpha}\exp[-A't^{\beta}] \,,
\end{align}
and
\begin{align}
\sigma_F(t) &\propto \sigma^\prime_{o}\sqrt{\left(1-\frac{B^\prime}{A^\prime \sigma^{\prime \, 2}_{o}}\right)\exp[-2A^\prime t^{\beta}] + \frac{B^\prime}{A^{\prime} \sigma^{\prime \, 2}_{o}}} \,.
\end{align}
This distribution is normalized also as in Eq.~(\ref{eq:norm_d}).

Observe that, if $A$ and $B$ are time independent, then the transformation $t^\beta \rightarrow \tau$ would transform the solution into a distribution formally equivalent to setting $\beta=1$. This case, which represents a common situation where the properties of the medium and its interaction with the dynamic system do not change with time, allows for a simple connection between the fractal properties and the parameter $q$, as will be shown in the next section.

\subsection{Fractal effects in the dynamics}

To study the fractal effects on the dynamics of a system diffusing in a fractal space it is necessary to establish a connection between the fractal quantities and the parameter $q$ in the PPE. It will be found that the dimensional gap $\delta d_f$, defined in Eq.~(\ref{eq:fractalgap}) is a relevant quantity to describe the dynamics.

By definition, the width of the distribution, $\Delta p_M$, is such that
\begin{equation}
    f(p_M+ \Delta p_M,t)=e^{-1/2}f(p_M,t)\,,
\end{equation}
which is the momentum for which the probability distribution is reduced by a factor $e^{-1/2}$. For the fractal distribution resulting form the FFPE, i.e. Eq.~(\ref{eq:f_fractal-2}), one has
\begin{equation}
\Delta p(t) = \sigma_F(t)^{1/\alpha} \,.
\end{equation}

The dynamical behaviour of the system governed by the PPE will be similar to the one controlled by the FFPE if the widths of both distributions are equal for any instant $t$, resulting in
\begin{equation}
\sigma(t) = \sigma_F(t)^{1/\alpha} \,.
\end{equation}
From here, one gets
%


\begin{align}
& \alpha  = \xi/2 =   1 - \delta d_f/2 \,, \label{eq:alpha} \\ & \beta = 1 \,,\label{eq:alpha_beta_d}
\end{align}
and in addition
\begin{align} 
A^\prime &= \alpha A = \left( 1 - \delta d_f/2 \right) A \,,  \label{eq:TransportCoeffA} \\
B^\prime &= \kappa A^\prime \sigma_o^{\prime \, 2} =  (2-q)  \left[ 1 -\delta d_f/2 \right] \left( \frac{2\pi \chi_q}{\lambda^2} \right)^{\frac{\delta d_f}{2}}  B \,, \label{eq:TransportCoeffB} \\
\sigma_o^\prime &= \sigma_o^\alpha \,.  
\end{align}

\section{Discussion and conclusions}

As an example of the application of the formalism developed above, this work will address the heavy-quark dynamics in the QGP, which is a problem of relevance in the understanding of the system formed at high-energy nuclear collisions.
This discussion will compare the fractal dynamics, as described by the distribution in Eq.~(\ref{eq:f_fractal-2}), with its continuous approximation, namely the solution of the PPE given by Eq.~(\ref{PPEsol-2}).

\begin{figure*}[t]
\centering
\includegraphics[width=0.40\textwidth]{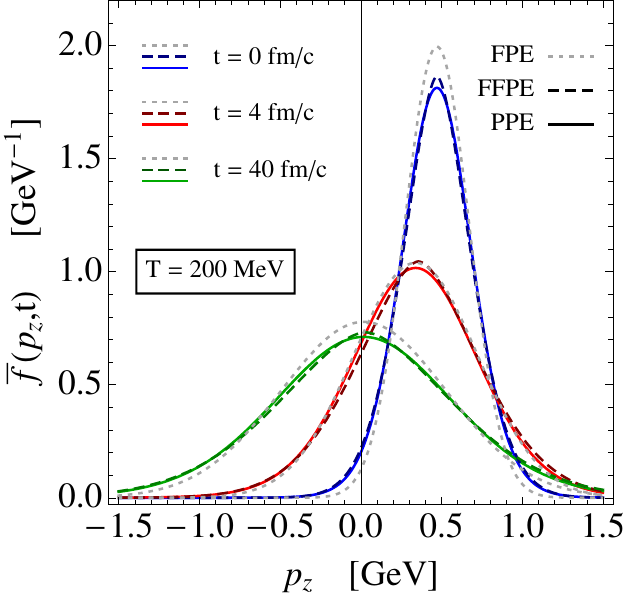}  
\hspace{1cm}  
\includegraphics[width=0.40\textwidth]{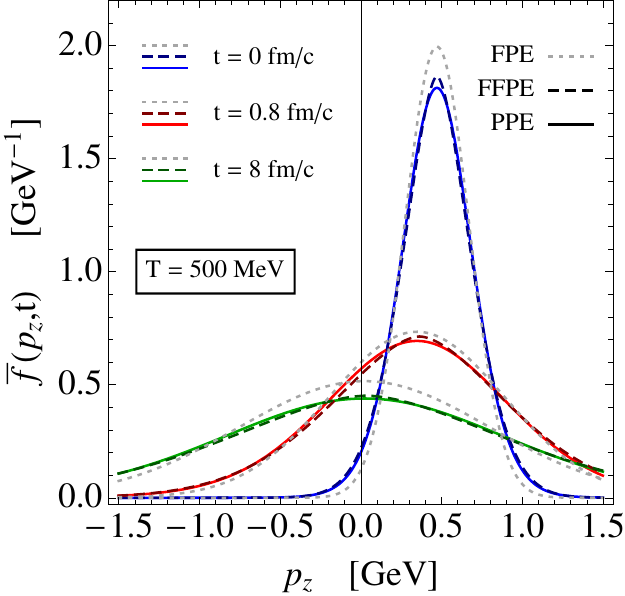}  
 \caption{Plots of the distribution functions integrated out in transverse momenta, $\bar f(p_z,t) \equiv \int_{-\infty}^\infty dp_x dp_y \bar f(\mathbf p,t)$ with $\bar f(\mathbf p,t) \equiv f(\mathbf p,t) / \lambda^3$, as a function of~$p_z$ for several values of $t$, for temperatures $T = 200$~MeV (left) and $T = 500$~MeV (right).  Solid and dashed lines correspond to $q = 1.10$, and dotted lines to $q = 1$. We have considered the rest mass of the heavy quark $m_0 = 1.5 \, \textrm{GeV}$, its initial momentum $p_o = 0.47\, \textrm{GeV}$ in the $z$-direction corresponding to the initial speed $v_o = 0.30\, c$, the initial width $\sigma_o = 0.2 \, \textrm{GeV}$, while $(A = 0.0820 \,\textrm{fm}^{-1}, B = 0.0216 \,\textrm{GeV}^2/\textrm{fm})$ for $T=200 \,\textrm{MeV}$ (left panel) and  $(A = 0.381 \,\textrm{fm}^{-1}, B = 0.228 \,\textrm{GeV}^2/\textrm{fm})$ for $T=500 \,\textrm{MeV}$ (right panel). We have considered $\lambda = 1 \, \textrm{GeV}$ for the multiplicative factor of the distribution function.}
\label{fig:fpz}
\end{figure*}

The numerical computation is performed for two different values of the medium temperature: $T=200\,\textrm{MeV}$ and $500\,\textrm{MeV}$. The value $q=1.10$ is adopted, following the discussion in Ref.~\cite{Megias:2022yrw}, and the normalization indicated in Eq.~(\ref{eq:norm_d}) is done with $\lambda = 1 \, \textrm{GeV}$. The embedding Euclidean space has dimension $d=3$. It is assumed that $\mathbf p_M(t) = p_M(t) \mathbf k$.  The values of the constants $A$ and $B$ have been taken from Ref.~\cite{Megias2023}, and are in agreement with Ref.~\cite{Svetitsky:1987gq}. For comparison with the standard FPE solution, the distributions were computed also with the value $q=1$.

With the conditions determined above for the calculation and from the formalism developed in the present work, it results 
\begin{align}
\alpha & = \frac{5 - 3 q}{2} \,, \\
\beta & = 1 \,, \\
A^\prime &= \alpha A = \frac{5-3q}{2} A \,, \\ 
B^\prime &= \frac{1}{2} (2-q) (5-3q) \left( \frac{2\pi \chi_q}{\lambda^2} \right)^{\frac{3}{2}(q-1)}  B \,, \\
\sigma_o^\prime &= \sigma_o^\alpha \,.
\end{align}
These equations establish the correct relationship between the transport coefficients in the PPE ($A$ and $B$) and the corresponding coefficients in the FFPE ($A'$ and $B'$), as well as the relationship between the initial values of the parameters $\sigma_o$ and $\sigma'_o$ for the PPE and FFPE, respectively.

Fig.~\ref{fig:fpz} shows the distributions obtained by using both the FFPE and the PPE for two distinct medium temperatures, namely, $T=200\,\textrm{MeV}$ (left panel) and $500\,\textrm{MeV}$ (right panel). The entropic index was set to $q=1.10$, and this choice is based on the relevant quark flavours involved in the dynamics, which depends on their masses~\cite{Megias2023}. The curves evidence a remarkable agreement of the solutions of the PPE and of the FFPE if the equations obtained in the present work are considered, especifically the equations between $q$ and the fractal dimension (Eq.~(\ref{eq:alpha-q-1D}) and Eq.~(\ref{eq:fractalgap})), and the equations for the correspondence among the transport coefficients (Eq.~(\ref{eq:TransportCoeffA}) and Eq.~(\ref{eq:TransportCoeffB})).

The agreement between the curves is remarkably good, which shows that the relationship between the transport coefficients is correct. These relationships result from the correct interpretation of the fractal aspects that are relevant for the dynamics of the system and the correct relationship between the entropic parameter $q$ and the fractal gap $\delta d_f$.

The dimension $d=\lceil d_f \rceil=3$ sets an upper limit for the fractal dimension in the heavy-quark dynamics.  Using $q=1.10$, Eq.~(\ref{eq:alpha}) gives $\delta d_f=0.3$, therefore $d_f=2.7$ is the fractal dimension of the momentum space. In the cases where the transport coefficients are time-independent, the fractal aspects of the time-space seem to be of limited effects in the dynamics, provided one takes into account the possibility of transforming $t \rightarrow \tau=t^{\beta}$, as discussed previously.

The distribution with $q=1$, which corresponds to the standard FPE,presents a similar shape, but its peak moves faster to stationary state position. This result is explained by the modification in the transport coefficient $A$ with the fractal dimension, as given by Eq.~(\ref{eq:TransportCoeffA}). The width of this distribution, however, remains always smaller than in the fractal cases. This is also a consequence of the fractality of the momentum space due to the modifications in the transport coefficient $B$, as described by Eq.~(\ref{eq:TransportCoeffB}).

This work has explored the interconnections between the traditional Fokker-Planck Equation (FPE) and its fractal counterpart, the Fractal Fokker-Planck Equation (FFPE), which is derived using the formalism of fractal derivatives. The study examined the solutions of both equations and compared their behaviours in the context of the dynamics of complex systems, particularly in the study of the quark-gluon plasma (QGP). The results evidence that the FFPE can provide a reasonable approximation for the dynamics of fractal systems while taking into account the non-local correlations present in such systems. The solutions of the FFPE exhibit behaviours that differ from those of the traditional FPE, which can be attributed to the fractal nature of the system.

The work performed here can interest researchers in areas such as the study of neutron stars~\cite{Annala:2019puf,Annala:2020rgx,Cardoso2017,Sen2023}, the study of strongly interacting particles, solar plasma~\cite{CORADDU2003473} and ionic diffusion~\cite{Curilef}, for instance. The micro-calculation of the transport coefficients for hadronic matter has been under debate~\cite{Sen2023}, with implications in hydrodynamical models. The results obtained here imply modifications in both the form of the solution for the dynamical equation and on the transport coefficients, and can be of interest to researchers in those fields.

\section{Acknowledgements}

A.D. is supported by the Project INCT-FNA (Instituto Nacional de Ci\^encia e Tecnologia - F\'{\i}sica Nuclear Aplicada) Proc. No. 464898/2014-5, by the Conselho Nacional de Desenvolvimento
Cient\'{\i}fico e Tecnol\'ogico (CNPq-Brazil), grant 304244/2018-0, by Project INCT- FNA Proc. No. 464 898/2014-5, and by FAPESP, Brazil grant 2016/17612-7. The work of E.M. is supported by the project PID2020-114767GB-I00 funded by MCIN/AEI/10.13039/501100011033, by the FEDER/Junta de Andaluc\'{\i}a-Consejer\'{\i}a de Econom\'{\i}a y Conocimiento 2014-2020 Operational Program under Grant A-FQM178-UGR18, and by Junta de Andaluc\'{\i}a under Grant FQM-225. The research of E.M. is also supported by the Ram\'on y Cajal Program of the Spanish MICIN under Grant RYC-2016-20678, and  by the ``Pr\'orrogas de Contratos Ram\'on y Cajal'' Program of the University of Granada.


\bibliographystyle{ieeetr}
\bibliography{FFPEvsPPE2}

\end{document}